\documentclass[aps,pre,onecolumn,superscriptaddress,nofootinbib]{revtex4-2}

\usepackage{amsmath,amssymb,bm,mathtools}
\usepackage{graphicx}
\usepackage{tikz}
\usetikzlibrary{arrows.meta,positioning,calc,fit,shapes.geometric}
\usepackage[colorlinks=true,linkcolor=blue,citecolor=blue,urlcolor=blue]{hyperref}
\hypersetup{
  pdftitle={Path-Dependent Energy Lagrangian for Irreversible Thermomechanical Systems},
  pdfauthor={Huilong Ren}
}

\newcommand{\GradX}{\operatorname{Grad}_{\!\mathbf X}}
\newcommand{\DivX}{\operatorname{Div}_{\!\mathbf X}}
\newcommand{\grad}{\operatorname{grad}}
\newcommand{\diver}{\operatorname{div}}
\newcommand{\sym}{\operatorname{sym}}
\newcommand{\tr}{\operatorname{tr}}

\begin{document}

\title{Path-Dependent Energy Lagrangian for Irreversible Thermomechanical Systems}
\author{Huilong Ren}
\email{hlren@tongji.edu.cn}
\affiliation{College of Civil Engineering, Tongji University, Shanghai 200092, China}
\date{November 4, 2025}

\begin{abstract}
We present a minimal Path-Dependent Energy Lagrangian (PDEL) that formulates, within a single action, the balance equations of mechanics and the entropy/heat equation for irreversible thermomechanical systems. The reversible part is the Helmholtz free energy, while irreversible effects enter through a history integral of channel powers. A single upper-limit/tangential variation rule makes the same instantaneous power appear as a dissipative force in the mechanical or internal-variable equations and as a positive source in the entropy/heat equation, closing the first law without double counting and guaranteeing nonnegative entropy production under mild monotonicity assumptions. The formulation preserves classical Lagrangian mechanics while accommodating standard dissipative models, including Kelvin--Voigt viscosity and diffusion, together with their associated heating, and clarifies the reversible character of thermomechanical cross terms. It offers a compact alternative to separate Rayleigh- or Onsager-type augmentations and to GENERIC/metriplectic brackets, with limited algebraic complexity and direct extension to multiphysics settings.
\end{abstract}

\maketitle

\section*{INTRODUCTION}

Classical Lagrangian mechanics generates conservative dynamics from a single action \cite{LandauLifshitz,Arnold}. By contrast, irreversible thermomechanical phenomena are commonly introduced as augmentations to the variational core through Rayleigh-type terms, linear force--flux closures, or bracket structures. In the rational approach, the first and second laws are imposed as separate local statements linked by constitutive restrictions and the entropy inequality \cite{TruesdellNoll,ColemanNoll,GurtinFriedAnand,TadmorMillerElliott}. Near equilibrium, Onsager reciprocity closes fluxes linearly with nonnegative quadratic production \cite{OnsagerI,OnsagerII}. Far from equilibrium, the Prigogine and de Groot--Mazur traditions and extended thermodynamics provide balance-form formulations and entropy-production principles \cite{Prigogine1947,Prigogine1962,deGrootMazur,Gyarmati}. A different line of development, GENERIC/metriplectic theory, represents conservation and production through compatible Poisson and metric operators \cite{GrmelaOttinger,Ottinger}. Energy--dissipation principles and gradient-flow theory recast evolution in terms of a free energy, a convex dissipation potential, and a metric structure \cite{Mielke,AmbrosioGigliSavare,Villani}. In mechanics and materials, irreversible effects are also formulated through microforce/microstress balances and variational treatments of inelasticity, viscosity, and phase transitions \cite{Maugin,RajagopalSrinivasa,SimoHughes,MieheEtAl,OrtizStainier}. Although these frameworks are effective, entropy production is generally treated outside the action that yields the mechanical equations, which can complicate power accounting when several couplings are present.

\begin{figure}[t]
\centering
\begin{tikzpicture}[
  >=Latex,
  every node/.style={font=\small},
  box/.style={draw,rounded corners=2pt,align=center,inner sep=3pt,minimum height=8mm},
  line/.style={draw,-{Latex[length=2.2mm]}}
]
\node[box,minimum width=42mm] (energy) at (0,1.65) {Classical route\\Energy balance (First Law)};
\node[box,minimum width=46mm] (entropy) at (-0.2,0.45) {Entropy inequality (Second Law)};
\node[box,minimum width=42mm] (closure) at (0,-0.85) {Constitutive/closure\\(compatibility checks)};
\node[box,minimum width=27mm] (pdel) at (5.2,1.65) {\textbf{PDEL}\\Single functional};
\node[box,minimum width=39mm] (mech) at (5.2,0.25) {Momentum $+$ internal\\variable equations};
\node[box,minimum width=36mm] (heat) at (5.2,-1.15) {Entropy/heat equation};
\draw[line] (energy.east) -- (pdel.west);
\draw[line] (entropy.east) -- ++(0.55,0) |- (pdel.west);
\draw[line] (closure.east) -- ++(1.05,0) |- (pdel.west);
\draw[line] (pdel.south) -- (mech.north);
\draw[line] (mech.south) -- (heat.north);
\end{tikzpicture}
\caption{Classical two-law route versus the single PDEL route. A single action, evaluated with the upper-limit/tangential variation rule, yields both the mechanical and entropy/heat equations.}
\label{fig:routes}
\end{figure}
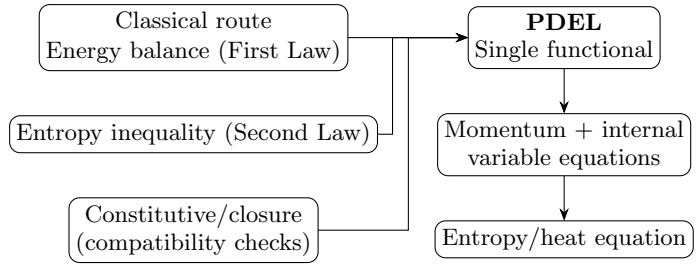

We propose a Path-Dependent Energy Lagrangian (PDEL): a single functional whose reversible part is a Helmholtz free energy and whose irreversible part is a history integral of instantaneous channel powers. A single upper-limit (tangential) variation---a reparameterization of time along the same path---makes the same channel power $\Phi_{\alpha}=Y_{\alpha}:\dot Z_{\alpha}\geq0$ appear with the two signs required by power balance: positive in the entropy/heat equation and negative when the corresponding channel equation is tested by its rate. The formulation therefore yields (i) the momentum and internal-variable equations and (ii) the entropy/heat equation, together with a channel-wise power audit that closes the first law without auxiliary multipliers or bracket structures. Figure~\ref{fig:routes} compares the conventional two-law route with the present construction.

The main features are as follows. (1) \emph{One functional for both laws.} Stationarity gives the entropy/heat equation directly; inserting $s=-\partial_{\theta}\phi$ gives the temperature form. Dissipative forces arise when the corresponding channel variables are varied, whereas the heat/entropy balance requires only the channel powers. (2) \emph{Channel-wise power balancing.} The same $\Phi_{\alpha}$ that produces a dissipative force $Y_{\alpha}$ also enters heat as $+\Phi_{\alpha}$, which gives the correct energy audit for each channel. (3) \emph{Minimal constitutive assumptions.} With $\mathbf K\succ0$ and nonnegative channel powers, the second law holds in production form; near equilibrium, Onsager closures are recovered through linear relations $Y_{\alpha}=L_{\alpha}\dot Z_{\alpha}$ \cite{OnsagerI,OnsagerII}. Convex dissipation potentials place the construction within the energy--dissipation/gradient-flow setting \cite{Mielke,AmbrosioGigliSavare}, while metric choices connect it with the GENERIC sector \cite{GrmelaOttinger,Ottinger}. (4) \emph{Direct multiphysics extension.} Diffusion, reactions, Joule heating, viscous stresses, and hereditary effects can be introduced as channels with powers $\Phi_{\alpha}$. Their sum then enters the heat/entropy equation with the appropriate sign, while variation with respect to the associated variables supplies the corresponding dissipative forces \cite{deGrootMazur,GurtinFriedAnand,Maugin,BirdStewartLightfoot}.

Unlike Rayleigh/Biot and contact/Herglotz-type nonconservative Lagrangians, which postulate a dissipation potential in rate space \cite{Biot,HalphenNguyen,Galley,Vermeeren,BravettiCruzTapias}, the present formulation integrates power in time, varies only its upper limit, and obtains the heat/entropy equation from the same action without doubled fields or contact algebra. This geometric modification keeps the entropy/heat source inside the action that produces the equations of motion and is consistent with the classical transport laws of Fourier and Fick \cite{deGrootMazur,BirdStewartLightfoot,Fourier,Fick} and with the rational continuum framework \cite{TruesdellNoll,ColemanNoll,GurtinFriedAnand}.

\section*{FRAMEWORK}

We work in a reference configuration $B_{0}$ with material coordinates $\mathbf X$, deformation $\boldsymbol\varphi(\mathbf X,t)$, displacement $\mathbf u=\boldsymbol\varphi-\mathbf X$, deformation gradient $\mathbf F=\GradX\boldsymbol\varphi$, Jacobian $J=\det\mathbf F$, and density $\rho=\rho_{0}/J$. The thermodynamic fields are temperature $\theta$ and entropy density $s$. Internal variables $\xi$ (plastic strain, damage, phase fraction, etc.) describe the material structure. Let $\phi(\mathbf F,\theta,\xi)$ be the Helmholtz free-energy density per reference volume.

\subsection*{Action and upper-limit/tangential variations}

The action is
\begin{equation}
S=\int_{t_{0}}^{T}\!\int_{B_{0}}
\left[\frac{\rho_{0}}{2}|\dot{\mathbf u}|^{2}
-\left(\phi(\mathbf F,\theta,\xi)+\theta s+\int_{t_{0}}^{t}D(\tau)\,\mathrm d\tau\right)\right]
\mathrm dV_{0}\,\mathrm dt,
\label{eq:action}
\end{equation}
with the instantaneous channel-power decomposition
\begin{equation}
D=D_{\mathrm{heat}}-\sum_{\alpha}\Phi_{\alpha},
\qquad
D_{\mathrm{heat}}=\DivX\mathbf q_{0},
\qquad
\mathbf q_{0}=-\mathbf K\GradX\theta,
\label{eq:D}
\end{equation}
where $\mathbf K\succ0$ and each nonthermal channel power $\Phi_{\alpha}\geq0$ (viscous, plastic, diffusive, reactive) is written as a force--rate pairing $\Phi_{\alpha}=Y_{\alpha}:\dot Z_{\alpha}$.

Four structural elements distinguish the action: (i) the explicit $+\theta s$ term, which enforces $s=-\partial_{\theta}\phi$ by stationarity; (ii) heat written as the divergence $\DivX\mathbf q_{0}$, which makes the heat flux explicit; (iii) modular nonnegative channel powers $\Phi_{\alpha}$ collected in $D$; and (iv) an upper-limit/tangential variation of $\int_{t_{0}}^{t}D\,\mathrm d\tau$, which places the instantaneous channel powers in the entropy balance.

\begin{figure}[t]
\centering
\begin{tikzpicture}[
  >=Latex,
  every node/.style={font=\small},
  box/.style={draw,rounded corners=2pt,align=center,inner sep=3pt,minimum height=8mm},
  line/.style={draw,-{Latex[length=2.2mm]}}
]
\node[box,minimum width=34mm] (channel) at (0,1.15) {Channel $\alpha$\\$\Phi_{\alpha}=Y_{\alpha}:\dot Z_{\alpha}\geq0$};
\node[box,minimum width=31mm] (heat) at (4.2,1.15) {Heat/entropy\\$+\Phi_{\alpha}/\theta$ in Eq.~\eqref{eq:entropyineq}};
\node[box,minimum width=35mm] (mech) at (0,-0.55) {Mechanics/structure\\$-Y_{\alpha}:\delta Z_{\alpha}$};
\node[box,minimum width=31mm] (cond) at (4.0,-0.55) {Conduction\\$D_{\mathrm{heat}}=\DivX\mathbf q_{0}$};
\node[box,minimum width=33mm] (boundary) at (7.6,-0.55) {Boundary heat flux\\$-\mathbf q_{0}\cdot\mathbf N$};
\draw[line] (channel.east) -- (heat.west);
\draw[line] (channel.south) -- (mech.north);
\draw[line] (cond.east) -- (boundary.west);
\end{tikzpicture}
\caption{Channel-wise power balancing. The same instantaneous power enters the heat/entropy equation as a positive source and the mechanical/structural energy with the opposite sign; conduction balances the boundary heat flux.}
\label{fig:power}
\end{figure}
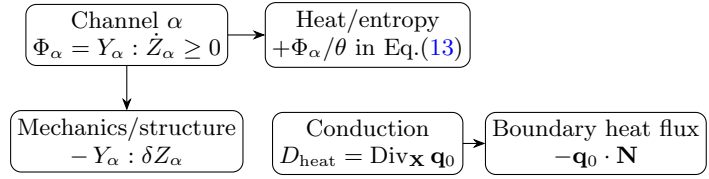

The upper-limit variation and tangential virtual variations are
\begin{equation}
\delta\!\left[\int_{t_{0}}^{t}D(\tau)\,\mathrm d\tau\right]
=D(t)\,\delta t,
\qquad
\delta Z_{\alpha}=\dot Z_{\alpha}\,\delta t,
\qquad
\delta s=\dot s\,\delta t.
\label{eq:tangential}
\end{equation}
Changing $t_{0}$ shifts $S$ by a time-independent constant and therefore leaves $\delta S$ unchanged. Reparameterize time along the same path as $t\mapsto t+\varepsilon\eta$, with $\eta(\cdot,t_{0})=0$, so that only the upper limit moves; $D$ is constitutive and is not varied internally (see Eqs.~\eqref{eq:upperlimit} and \eqref{eq:dragged}). Leibniz's rule and dominated convergence give $\delta\int_{t_{0}}^{t}D\,\mathrm d\tau=D(t)\delta t$. Since only the time label slides, $\delta s=\dot s\delta t$ and $\delta Z_{\alpha}=\dot Z_{\alpha}\delta t$. The identity $(Y_{\alpha}:\dot Z_{\alpha})\delta t=Y_{\alpha}:\delta Z_{\alpha}$ produces the dissipative force while retaining the same power in the entropy/heat equation, as illustrated in Fig.~\ref{fig:power}.

\subsection*{Conjugacy and cross derivatives}

Variation with respect to $\theta$ gives the temperature--entropy conjugacy
\begin{equation}
s=-\partial_{\theta}\phi(\mathbf F,\theta,\xi).
\label{eq:conjugacy}
\end{equation}
Accordingly, define the specific heat per unit reference mass by
$\rho_{0}c(\theta)=-\theta\,\partial_{\theta\theta}\phi\geq0$.
The cross derivatives $\partial_{\theta\mathbf F}\phi$ and $\partial_{\theta\xi}\phi$ describe reversible thermomechanical and thermostructural couplings. They appear as conversion terms in the temperature equation but do not contribute to entropy production.

\subsection*{Stationarity: balances and boundary conditions}

Varying $\mathbf u$, $s$, and each channel variable $Z_{\alpha}$ in Eq.~\eqref{eq:action}, subject to Eq.~\eqref{eq:tangential}, gives the field equations and natural boundary conditions.

\emph{Momentum (reference form).}
The conservative stress is $\mathbf P=\partial_{\mathbf F}\phi(\mathbf F,\theta,\xi)$. A mechanical channel with $Z_{\alpha}$ kinematically tied to $\mathbf u$ produces a dissipative stress $\bm\sigma_{\alpha}^{(X)}$ such that $Y_{\alpha}:\delta Z_{\alpha}=\bm\sigma_{\alpha}^{(X)}:\GradX\delta\mathbf u$ before spatial integration by parts. The balance and traction condition are
\begin{equation}
\rho_{0}\ddot{\mathbf u}
-\DivX\!\left(\mathbf P+\sum_{\alpha\in\mathrm{mech}}\bm\sigma_{\alpha}^{(X)}\right)
=\mathbf b_{0},
\qquad
\left(\mathbf P+\sum_{\alpha\in\mathrm{mech}}\bm\sigma_{\alpha}^{(X)}\right)\mathbf N
=\mathbf t_{0}
\quad\text{on }\partial B_{0}.
\label{eq:momentum}
\end{equation}

\emph{Internal variables.}
For structural channels, stationarity gives a dissipative microforce $\pi_{\alpha}$ together with the conservative driving force $\partial_{\xi}\phi$:
\begin{equation}
\partial_{\xi}\phi(\mathbf F,\theta,\xi)
+\sum_{\alpha\in\xi}\pi_{\alpha}=0.
\label{eq:internal}
\end{equation}

\emph{Entropy/heat.}
Entropy variation gives, with volumetric heating $r_{0}$,
\begin{equation}
\theta\dot s+\DivX\mathbf q_{0}-\sum_{\alpha}\Phi_{\alpha}=r_{0},
\qquad
\mathbf q_{0}\cdot\mathbf N=\bar q
\quad\text{on }\partial B_{0}.
\label{eq:entropyheat}
\end{equation}
Thus, the same nonthermal powers $\Phi_{\alpha}$ that generate dissipative forces also enter as positive heat sources in the temperature form below. Fourier or non-Fourier heat-flux laws can be introduced without changing these identities.

\subsection*{Temperature form and reversible cross terms}

Substituting Eq.~\eqref{eq:conjugacy} into Eq.~\eqref{eq:entropyheat} and using $\mathbf q_{0}=-\mathbf K\GradX\theta$ gives
\begin{equation}
\rho_{0}c(\theta)\dot\theta
=\DivX(\mathbf K\GradX\theta)
+\sum_{\alpha}\Phi_{\alpha}
+\theta\,\partial_{\theta\mathbf F}\phi:\dot{\mathbf F}
+\theta\,\partial_{\theta\xi}\phi\,\dot\xi
+r_{0}.
\label{eq:temperature}
\end{equation}
The cross terms are reversible, sign-indefinite transfer terms and therefore do not enter the entropy production.

\section*{THERMODYNAMIC CONSISTENCY}

Define the internal-energy density $e:=\phi+\theta s$. Using Eq.~\eqref{eq:conjugacy},
\begin{equation}
\dot e=\mathbf P:\dot{\mathbf F}+\theta\dot s+\partial_{\xi}\phi\,\dot\xi.
\label{eq:edot}
\end{equation}
Equation~\eqref{eq:entropyheat} then gives the local internal-energy identity
\begin{equation}
\dot e
=\mathbf P:\dot{\mathbf F}-\DivX\mathbf q_{0}+r_{0}
+\partial_{\xi}\phi\,\dot\xi+\sum_{\alpha}\Phi_{\alpha}.
\label{eq:localenergy}
\end{equation}
Testing the complete mechanical and structural equations by their actual rates, integrating over $B_{0}$, and adding Eq.~\eqref{eq:localenergy} gives the global first law for the closed internal channels considered here:
\begin{align}
\frac{\mathrm d}{\mathrm dt}\int_{B_{0}}
\left(\frac{\rho_{0}}{2}|\dot{\mathbf u}|^{2}+e\right)\mathrm dV_{0}
={}&\int_{\partial B_{0}}
\left(\mathbf P+\sum_{\alpha\in\mathrm{mech}}\bm\sigma_{\alpha}^{(X)}\right)\mathbf N\cdot\dot{\mathbf u}\,\mathrm dA_{0}
+\int_{B_{0}}\mathbf b_{0}\cdot\dot{\mathbf u}\,\mathrm dV_{0}
\nonumber\\
&-\int_{\partial B_{0}}\mathbf q_{0}\cdot\mathbf N\,\mathrm dA_{0}
+\int_{B_{0}}r_{0}\,\mathrm dV_{0}.
\label{eq:globalenergy}
\end{align}
The channel powers do not appear as external sources in Eq.~\eqref{eq:globalenergy}; they cancel against the corresponding mechanical or structural power through the channel equations.

\paragraph*{Lemma 1 (Power-balancing projection).}
Let $D=\DivX\mathbf q_{0}-\sum_{\alpha}\Phi_{\alpha}$, with $\Phi_{\alpha}=Y_{\alpha}:\dot Z_{\alpha}$. Under the upper-limit/tangential variation rule in Eq.~\eqref{eq:tangential}, the same $\Phi_{\alpha}$ appears as $+\Phi_{\alpha}$ in Eq.~\eqref{eq:entropyheat} and as the force $+Y_{\alpha}$ in the $Z_{\alpha}$-equation. Writing the conservative part of that equation as $\mathcal E_{\alpha}$, testing $\mathcal E_{\alpha}+Y_{\alpha}=0$ by $\dot Z_{\alpha}$ gives, channel-wise,
\begin{equation}
\Phi_{\alpha}+\mathcal E_{\alpha}:\dot Z_{\alpha}=0,
\label{eq:projection}
\end{equation}
which proves the cancellation used in Eq.~\eqref{eq:globalenergy}.

Dividing Eq.~\eqref{eq:entropyheat} by $\theta>0$ and using
$\theta^{-1}\DivX\mathbf q_{0}=\DivX(\mathbf q_{0}/\theta)+\mathbf q_{0}\cdot\GradX\theta/\theta^{2}$
gives the local entropy-production inequality
\begin{equation}
\dot s+\DivX\!\left(\frac{\mathbf q_{0}}{\theta}\right)-\frac{r_{0}}{\theta}
=\frac{\GradX\theta\cdot\mathbf K\GradX\theta}{\theta^{2}}
+\sum_{\alpha}\frac{\Phi_{\alpha}}{\theta}
\geq0,
\label{eq:entropyineq}
\end{equation}
provided $\mathbf K\succ0$ and each channel is monotone, $\Phi_{\alpha}\geq0$. Here $r_{0}/\theta$ is the external entropy supply and is therefore separated from the internal production.

\paragraph*{Proposition 1 (Cross terms produce no entropy).}
The reversible cross terms in Eq.~\eqref{eq:temperature}, $\theta\partial_{\theta\mathbf F}\phi:\dot{\mathbf F}$ and $\theta\partial_{\theta\xi}\phi\dot\xi$, do not appear in Eq.~\eqref{eq:entropyineq}. They therefore cannot replace $\sum_{\alpha}\Phi_{\alpha}$ as sources of entropy production. Instead, they describe reversible, sign-indefinite energy exchange between thermal and nonthermal storage.

\emph{Assumptions.}
The minimal conditions are $c(\theta)\geq0$, $\mathbf K\succ0$, and channel monotonicity $\Phi_{\alpha}\geq0$. Frame indifference is enforced through objective choices of $\phi$ and of the channel pairings; for example, $\Phi_{\mathrm{KV}}=2\eta\,\mathbf D:\mathbf D+\lambda(\diver\mathbf v)^{2}$ with $\mathbf D=\sym\grad\mathbf v$.

\section*{COMPACT EXAMPLES}

\subsection*{Compressible thermomechanics (ideal gas)}

Let $a(\rho,\theta)=c_{v}\theta[1-\ln(\theta/\theta_{*})]+R_{s}\theta\ln(\rho/\rho_{*})$ be the free energy per unit mass, and set $\phi=\rho_{0}a(\rho,\theta)$ with $\rho=\rho_{0}/J$. Then
\begin{equation}
\frac{s}{\rho_{0}}=c_{v}\ln(\theta/\theta_{*})-R_{s}\ln(\rho/\rho_{*}),
\qquad
p=\rho^{2}\partial_{\rho}a=R_{s}\rho\theta,
\label{eq:idealgas}
\end{equation}
which gives the Cauchy stress $\bm\sigma=-p\mathbf I$ and, in the current configuration,
$\rho c_{v}\dot\theta=-p\diver\mathbf v+\diver(k\grad\theta)+\sum_{\alpha}\Phi_{\alpha}^{(\mathrm{sp})}+r$ \cite{GurtinFriedAnand,TadmorMillerElliott}.

\subsection*{Kelvin--Voigt viscous solid}

In a small-strain Kelvin--Voigt specialization, define the channel power
\begin{equation}
\Phi_{\mathrm{KV}}
=\bm\sigma_{\mathrm{visc}}^{(X)}:\GradX\dot{\mathbf u},
\qquad
\bm\sigma_{\mathrm{visc}}^{(X)}
=2\eta\,\sym(\GradX\dot{\mathbf u})
+\lambda\,\tr(\GradX\dot{\mathbf u})\mathbf I,
\label{eq:KVpower}
\end{equation}
with $\eta,\lambda\geq0$. The upper-limit/tangential variation gives the viscous stress in momentum and the same power in heat:
\begin{align}
\rho_{0}\ddot{\mathbf u}-\DivX\!\left(\mathbf P+\bm\sigma_{\mathrm{visc}}^{(X)}\right)
&=\mathbf b_{0},
\qquad
\left(\mathbf P+\bm\sigma_{\mathrm{visc}}^{(X)}\right)\mathbf N=\mathbf t_{0},
\label{eq:KVmomentum}\\
\rho_{0}c(\theta)\dot\theta
&=\DivX(\mathbf K\GradX\theta)+\Phi_{\mathrm{KV}}
+\theta\partial_{\theta\mathbf F}\phi:\dot{\mathbf F}
+\theta\partial_{\theta\xi}\phi\dot\xi+r_{0}.
\label{eq:KVtemperature}
\end{align}
In the spatial frame, $\Phi_{\mathrm{KV}}=2\eta\,\mathbf D:\mathbf D+\lambda(\diver\mathbf v)^{2}\geq0$, which ensures nonnegative viscous entropy production \cite{GurtinFriedAnand,RajagopalSrinivasa,SimoHughes}.

\subsection*{Diffusion channel}

Let the concentration $c$ be a field with chemical potential $\mu=\partial_{c}\phi(c,\theta)$ and flux $\mathbf j=-M\GradX\mu$ ($M>0$). The corresponding channel power is
\begin{equation*}
\Phi_{\mathrm{diff}}=-\mathbf j\cdot\GradX\mu
=M|\GradX\mu|^{2}\geq0.
\end{equation*}
It enters the heat equation as $+\Phi_{\mathrm{diff}}$ and contributes $\Phi_{\mathrm{diff}}/\theta$ to the entropy production \cite{Mielke}.

\subsection*{Electrochemistry (battery half-cell)}

Let $c$ be the electrolyte concentration and $\varphi$ the electric potential. Take $\mu=\partial_{c}\phi(c,\theta)$, diffusive flux $\mathbf j=-M(c,\theta)\nabla_{X}\mu$, current $\mathbf i=-\sigma(c,\theta)\nabla_{X}\varphi$, and interfacial reaction rate $R=LA$ near equilibrium, with affinity $A$. The channel powers are
$\Phi_{\mathrm{diff}}=-\mathbf j\cdot\nabla_{X}\mu=M|\nabla_{X}\mu|^{2}$,
$\Phi_{\mathrm{Joule}}=-\mathbf i\cdot\nabla_{X}\varphi=\sigma|\nabla_{X}\varphi|^{2}$, and
$\Phi_{\mathrm{rxn}}=AR\geq0$. The temperature equation is
\begin{equation*}
\rho_{0}c(\theta)\dot\theta
=\operatorname{Div}_{X}(K\nabla_{X}\theta)
+\Phi_{\mathrm{diff}}+\Phi_{\mathrm{Joule}}+\Phi_{\mathrm{rxn}}
+\theta\partial_{\theta\xi}\phi\dot\xi+r_{0}.
\end{equation*}
Together with the species balance $\dot c+\operatorname{Div}_{X}\mathbf j=\pm R$ and charge balance $\operatorname{Div}_{X}\mathbf i=0$, this example shows how the nonthermal channel powers enter heat with the required sign while the corresponding dissipative fluxes and forces are retained in the field equations.

\section*{RELATION TO EXISTING FRAMEWORKS AND OUTLOOK}

Rayleigh/d'Alembert formulations add dissipative forces but require a separate thermal balance. Onsager and GENERIC/metriplectic formulations unify energy and entropy through additional algebraic structure \cite{OnsagerI,GrmelaOttinger,Ottinger}. The present construction instead uses a single path term evaluated by the upper-limit/tangential rule. The same channel power enters heat as $+\Phi$ and the mechanical or structural power with the opposite sign, without auxiliary multipliers or bracket structures.

\emph{Outlook.}
The formulation extends directly to multiphysics settings: diffusion, electromagnetics, and chemistry enter as additional state variables and channel powers in $D$, with the same power-balancing projection closing the laws. History effects can be incorporated by representing channel powers through hereditary integrals. These extensions preserve the same minimal structure \cite{Mielke,GrmelaOttinger,Ottinger}.

\begin{acknowledgments}
The author acknowledges support from the Fundamental Research Funds for the Central Universities under Grant No.~02302350113.
\end{acknowledgments}

\section*{END MATTER}

\subsection*{A. Tangential variations of path variables: upper-limit rule}

Let $B_{0}$ be the reference body, $t\in[t_{0},T]$, and
\begin{equation*}
\mathcal P(\mathbf X,t):=\int_{t_{0}}^{t}D(\mathbf X,\tau)\,\mathrm d\tau,
\qquad
D(\cdot,\cdot)\in L^{1}\!\left((t_{0},T);L^{1}(B_{0})\right).
\end{equation*}
Assume that the state $(\theta,s,\{Z_{\alpha}\},\ldots)$ is absolutely continuous in time with values in a Hilbert space (Bochner $AC$); in practice, $AC([t_{0},T];L^{2}(B_{0}))$ is sufficient, consistent with standard chain rules in continuum thermodynamics \cite{GurtinFriedAnand,TruesdellNoll,ColemanNoll}.

\emph{Time reparameterization.}
For a smooth test function $\eta(\mathbf X,t)$ with $\eta(\cdot,t_{0})=0$ (and, optionally, $\eta(\cdot,T)=0$), consider the near-identity reparameterization
\begin{equation*}
t_{\varepsilon}(\mathbf X,t)=t+\varepsilon\eta(\mathbf X,t),
\qquad
\delta t:=\eta.
\end{equation*}
By Leibniz's rule and dominated convergence, using $D\in L^{1}$, the Gâteaux derivative of the path integral at $\varepsilon=0$ is
\begin{equation}
\left.\frac{\mathrm d}{\mathrm d\varepsilon}\right|_{\varepsilon=0}
\mathcal P(\mathbf X,t_{\varepsilon})
=D(\mathbf X,t)\eta(\mathbf X,t)
\quad\Longrightarrow\quad
\delta\int_{t_{0}}^{t}D(\mathbf X,\tau)\,\mathrm d\tau
=D(\mathbf X,t)\delta t.
\label{eq:upperlimit}
\end{equation}
The path is not perturbed; only the upper limit is varied. Thus, no internal variation of $D$ is taken in this operation.

\emph{Tangential variations in state space.}
Because the time label slides along the same path, any dragged field $w(\mathbf X,t)$ satisfies
\begin{equation}
\delta w(\mathbf X,t)=\dot w(\mathbf X,t)\delta t,
\qquad
\text{in particular}\quad
\delta s=\dot s\delta t,
\quad
\delta Z_{\alpha}=\dot Z_{\alpha}\delta t.
\label{eq:dragged}
\end{equation}
Combining Eqs.~\eqref{eq:upperlimit} and \eqref{eq:dragged} gives the power identity
\begin{equation}
\big(Y_{\alpha}:\dot Z_{\alpha}\big)\delta t
=Y_{\alpha}:\delta Z_{\alpha},
\label{eq:poweridentity}
\end{equation}
which is the algebraic relation by which a channel power contributes both to heat through $\delta t$ and to the corresponding force through $\delta Z_{\alpha}$.

\emph{Weak/distributional form.}
If $D\in L^{1}$ and the state is only $L^{2}$ in space, Eqs.~\eqref{eq:upperlimit}--\eqref{eq:dragged} hold in the distributional sense on $B_{0}\times(t_{0},T)$. Test functions may be taken in $H^{1}(t_{0},T;L^{2}(B_{0}))\cap L^{2}(t_{0},T;H^{1}(B_{0}))$, consistently with standard assumptions in continuum thermodynamics and gradient-flow analysis \cite{GurtinFriedAnand,Mielke}.

\emph{Remark (contrast with dissipation potentials).}
The rule above differs from Rayleigh/Biot or contact/Herglotz constructions that vary a rate potential directly \cite{Biot,HalphenNguyen}. Here, power is integrated in time and only its upper limit is varied; the channel force $Y_{\alpha}$ appears through Eq.~\eqref{eq:poweridentity}, while the same power enters heat through the $\delta t$ contribution.

\subsection*{B. Channel equations and the power-balancing projection}

\emph{Channel equations from stationarity.}
Let the action be
\begin{equation*}
S=\int_{t_{0}}^{T}\!\int_{B_{0}}
\left[\mathcal K-\left(\phi(\cdot)+\theta s+\int_{t_{0}}^{t}D\,\mathrm d\tau\right)\right]
\mathrm dV_{0}\,\mathrm dt,
\end{equation*}
where $\mathcal K$ is the kinetic term, if present, and
$D=\DivX\mathbf q_{0}-\sum_{\alpha}\Phi_{\alpha}$ with
$\Phi_{\alpha}=Y_{\alpha}:\dot Z_{\alpha}\geq0$.
Using Eq.~\eqref{eq:poweridentity}, the variation of the path term contributes two parts: (i) the heat/entropy term $-D\delta t$ from the upper limit and (ii) the force term $-Y_{\alpha}:\delta Z_{\alpha}$ for each channel. After the usual spatial integrations by parts, when $Z_{\alpha}$ is kinematically tied to fields with gradients, stationarity gives:
\begin{itemize}
\item \emph{Entropy/heat balance:}
$\theta\dot s+\DivX\mathbf q_{0}-\sum_{\alpha}\Phi_{\alpha}=r_{0}$, with $\mathbf q_{0}=-\mathbf K\GradX\theta$ and $\mathbf K\succ0$.
\item \emph{Channel equations:}
$\mathcal E_{\alpha}(Z_{\alpha},\text{state})+Y_{\alpha}=0$ in $B_{0}$, where $\mathcal E_{\alpha}$ collects the conservative or microforce contribution from the reversible sector, such as gradients of internal variables or elastic couplings \cite{GurtinFriedAnand,Maugin,SimoHughes}.
\end{itemize}

\emph{Power-balancing projection (local identity).}
Testing the channel equation by the actual channel rate $\dot Z_{\alpha}$ gives
\begin{equation*}
(\mathcal E_{\alpha}+Y_{\alpha}):\dot Z_{\alpha}=0
\quad\Longrightarrow\quad
\Phi_{\alpha}+\mathcal E_{\alpha}:\dot Z_{\alpha}=0.
\end{equation*}
Adding this identity for every $\alpha$ to the entropy/heat equation yields the local first-law statement: each nonthermal channel power $\Phi_{\alpha}$ appears once as a positive heat source and once with the opposite sign in the mechanical or structural power through $\mathcal E_{\alpha}:\dot Z_{\alpha}$. This is the no-double-counting property.


\begin{thebibliography}{30}

\bibitem{LandauLifshitz}
L. D. Landau and E. M. Lifshitz, \emph{Mechanics}, Vol.~1 (Pergamon Press, Oxford, 1960).

\bibitem{Arnold}
V. I. Arnol'd, \emph{Mathematical Methods of Classical Mechanics}, Vol.~60 (Springer, New York, 2013).

\bibitem{TruesdellNoll}
C. Truesdell and W. Noll, \emph{The Non-Linear Field Theories of Mechanics}, in \emph{Handbuch der Physik}, Vol.~III/3 (Springer, Berlin, 1965).

\bibitem{ColemanNoll}
B. D. Coleman and W. Noll, Arch. Ration. Mech. Anal. \textbf{13}, 167 (1963).

\bibitem{GurtinFriedAnand}
M. E. Gurtin, E. Fried, and L. Anand, \emph{The Mechanics and Thermodynamics of Continua} (Cambridge University Press, Cambridge, 2010).

\bibitem{TadmorMillerElliott}
E. B. Tadmor, R. E. Miller, and R. S. Elliott, \emph{Continuum Mechanics and Thermodynamics: From Fundamental Concepts to Governing Equations} (Cambridge University Press, Cambridge, 2012).

\bibitem{OnsagerI}
L. Onsager, Phys. Rev. \textbf{37}, 405 (1931).

\bibitem{OnsagerII}
L. Onsager, Phys. Rev. \textbf{38}, 2265 (1931).

\bibitem{Prigogine1947}
I. Prigogine, \emph{Étude thermodynamique des phénomènes irréversibles} (Desoer, Liège, 1947).

\bibitem{Prigogine1962}
I. Prigogine, \emph{Introduction to Thermodynamics of Irreversible Processes} (Interscience, New York, 1962).

\bibitem{deGrootMazur}
S. R. de Groot and P. Mazur, \emph{Non-Equilibrium Thermodynamics} (North-Holland, Amsterdam, 1962).

\bibitem{Gyarmati}
I. Gyarmati, \emph{Non-Equilibrium Thermodynamics: Field Theory and Variational Principles} (Springer, Berlin, 1970).

\bibitem{GrmelaOttinger}
M. Grmela and H. C. \"Ottinger, Phys. Rev. E \textbf{56}, 6620 (1997).

\bibitem{Ottinger}
H. C. \"Ottinger, \emph{Beyond Equilibrium Thermodynamics} (Wiley, Hoboken, 2005).

\bibitem{Mielke}
A. Mielke, Nonlinearity \textbf{24}, 1329 (2011).

\bibitem{AmbrosioGigliSavare}
L. Ambrosio, N. Gigli, and G. Savaré, \emph{Gradient Flows: In Metric Spaces and in the Space of Probability Measures} (Springer, Basel, 2005).

\bibitem{Villani}
C. Villani, \emph{Optimal Transport: Old and New}, Vol.~338 (Springer, Berlin, 2009).

\bibitem{Maugin}
G. A. Maugin, \emph{The Thermomechanics of Nonlinear Irreversible Behaviors}, Vol.~27 (World Scientific, Singapore, 1999).

\bibitem{RajagopalSrinivasa}
K. R. Rajagopal and A. R. Srinivasa, Proc. R. Soc. Lond. A \textbf{460}, 631 (2004).

\bibitem{SimoHughes}
J. C. Simo and T. J. R. Hughes, \emph{Computational Inelasticity} (Springer, New York, 1998).

\bibitem{MieheEtAl}
C. Miehe, M. Hofacker, and F. Welschinger, Comput. Methods Appl. Mech. Eng. \textbf{199}, 2765 (2010).

\bibitem{OrtizStainier}
M. Ortiz and L. Stainier, Comput. Methods Appl. Mech. Eng. \textbf{171}, 419 (1999).

\bibitem{BirdStewartLightfoot}
R. B. Bird, W. E. Stewart, and E. N. Lightfoot, \emph{Transport Phenomena} (Wiley, New York, 1960).

\bibitem{Biot}
M. A. Biot, Phys. Rev. \textbf{97}, 1463 (1955).

\bibitem{HalphenNguyen}
B. Halphen and Q. S. Nguyen, J. Mécanique \textbf{14}, 39 (1975).

\bibitem{Galley}
C. R. Galley, Phys. Rev. Lett. \textbf{110}, 174301 (2013).

\bibitem{Vermeeren}
M. Vermeeren, A. Bravetti, and M. Seri, J. Phys. A: Math. Theor. \textbf{52}, 445206 (2019).

\bibitem{BravettiCruzTapias}
A. Bravetti, H. Cruz, and D. Tapias, Ann. Phys. \textbf{376}, 17 (2017).

\bibitem{Fourier}
J. B. J. Fourier, \emph{Théorie analytique de la chaleur} (Gauthier-Villars, Paris, 1888).

\bibitem{Fick}
A. Fick, Philos. Mag. \textbf{10}, 30 (1855).

\end{thebibliography}
\end{document}